\documentclass[conference]{IEEEtran}
\IEEEoverridecommandlockouts
\usepackage{cite}
\usepackage{amsmath,amssymb,amsfonts}
\usepackage{algorithmic}
\usepackage{graphicx}
\usepackage{textcomp}
\usepackage{xcolor}
\usepackage[ruled,linesnumbered,nofillcomment]{algorithm2e}
\usepackage{pdfpages}
\def\BibTeX{{\rm B\kern-.05em{\sc i\kern-.025em b}\kern-.08em
    T\kern-.1667em\lower.7ex\hbox{E}\kern-.125emX}}
\begin{document}
\null%
\includepdf[pages={1}]{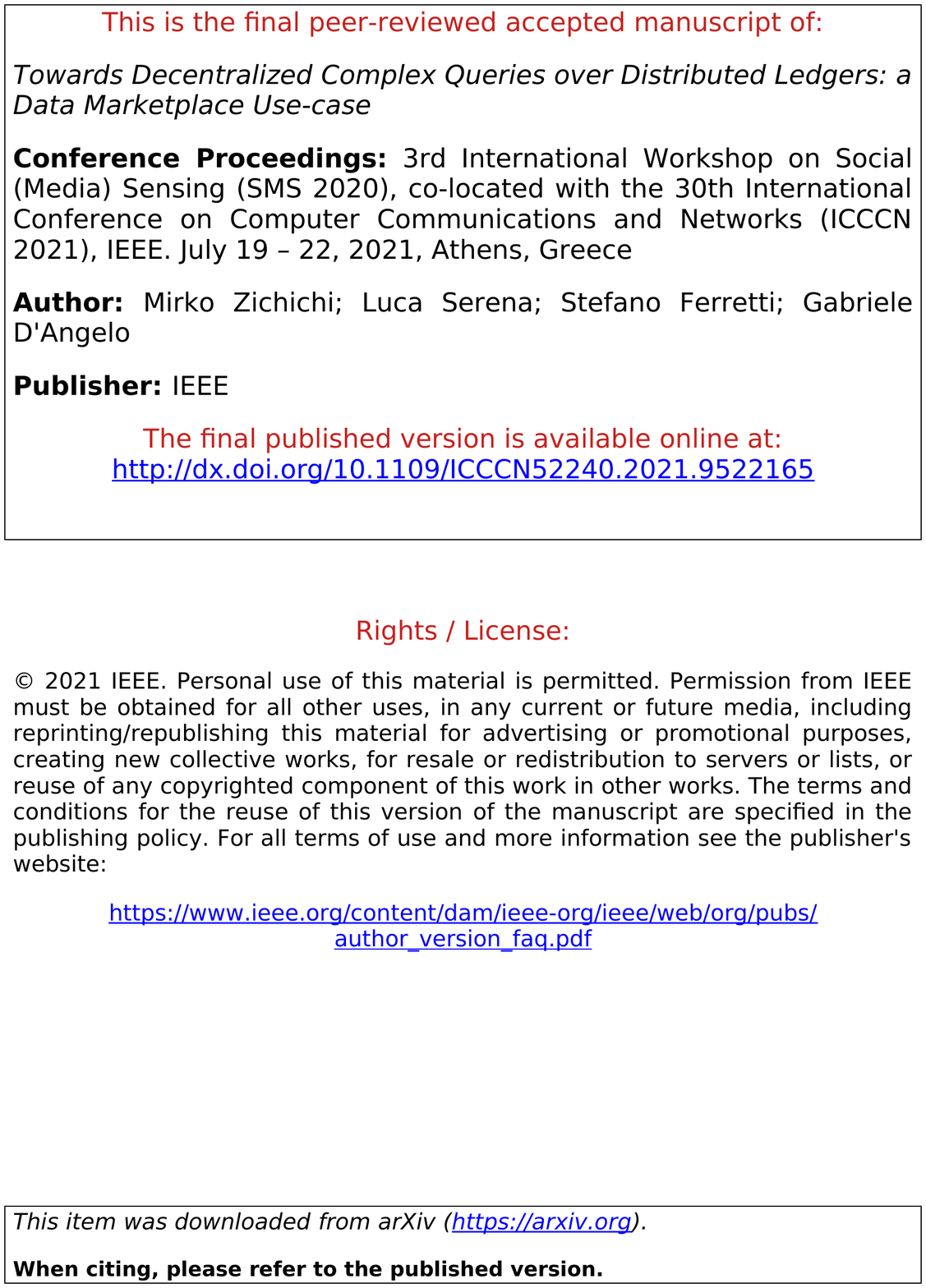}

\title{Towards Decentralized Complex Queries over Distributed Ledgers: a Data Marketplace Use-case
%Keyword-based Search in Decentralized\\ Data Marketplaces
\thanks{This work has received funding from the European Union’s Horizon 2020 research and innovation programme under the Marie Skłodowska-Curie International Training Network European Joint Doctorate grant agreement No 814177 Law, Science and Technology Joint Doctorate - Rights of Internet of Everything.}
}

\author{\IEEEauthorblockN{
Mirko Zichichi\IEEEauthorrefmark{1}\IEEEauthorrefmark{4},
Luca Serena\IEEEauthorrefmark{2},
Stefano Ferretti\IEEEauthorrefmark{3},
Gabriele D'Angelo\IEEEauthorrefmark{4}
}
\IEEEauthorblockA{\IEEEauthorrefmark{1}Ontology Engineering Group, Universidad Politécnica de Madrid, Spain}
\IEEEauthorblockA{\IEEEauthorrefmark{2}CIRI ICT, University of Bologna, Italy}
\IEEEauthorblockA{\IEEEauthorrefmark{3}Department of Pure and Applied Sciences, University of Urbino ``Carlo Bo", Italy}
\IEEEauthorblockA{\IEEEauthorrefmark{4}Department of Computer Science and Engineering, University of Bologna, Italy}
\emph{mirko.zichichi@upm.es, luca.serena2@unibo.it, stefano.ferretti@uniurb.it, g.dangelo@unibo.it} 
}

\maketitle

\begin{abstract}
Distributed Ledger Technologies (DLT) and Decentralized File Storages (DFS) are becoming increasingly used to create common, decentralized and trustless infrastructures where participants interact and collaborate in Peer-to-Peer interactions.
A prominent use case is represented by decentralized data marketplaces, where users are consumers and providers at the same time, and trustless interactions are required. However, data in DLTs and DFS are usually unstructured and there are no efficient mechanisms to query a certain type of data for the search in the market. 
In this paper, we propose the use of a Distributed Hash Table (DHT) as a layer on top of DLTs where, once the data are acquired and stored in the ledger, these can be searched through multiple keyword based queries, thanks to the lookup functionalities offered by the DHT. The DHT network is a hypercube overlay structure, organized for an efficient processing of multiple keyword-based queries.
We provide the architecture of such solution for a decentralized data marketplace and an analysis based on a simulation that proves the viability of the proposed approach.
\end{abstract}


%%%%%%%%%%%%%%%%%%%%%%%%%%%%%%%%%%%%%%%%%%%%%%%%%%%%%%%%%%%%%%%%%%%%%%%%%%%%%%%%%%%%%%%%%%%%%%%%%%%%%%%%%%%%%%%%%%%%%%%%%%%%%%%%%%%%%%%%%%%%%%%%%%%%%%%%%%%%%%%%%%%%%%%%%%%%%%%%%%%%%%%%%%%%%%%%%%%%%%%%%%%%%%%%%%%%%%%%%%%%%%%%%
\section{Introduction}
The transformation brought about by digital technologies has data at its core and has had a significant impact on economies and societies around the world. 
%Making more data available and improving the way that data is used is therefore a key challenge, as a small number of Big Tech companies currently own much of the world's data. 
The ability to easily get hold of data has the potential to create a data market where more and more users are consumers and providers at the same time. 
However, obtaining large amounts of data 
%to work with is not an easy task. The reason for this is that, besides having to collect large quantities of meaningful data for one's own purposes, obtaining data 
that is not of dubious or false origin is often a challenge. 

In order to tackle this issue, Distributed Ledger Technologies (DLT) and the realm of decentralized technologies (e.g. Decentralized File Storages (DFS)), that are emerging around them, come to the rescue~\cite{8656952,zichichi2020framework}. By creating a common, decentralized and trustless infrastructure, i.e. a decentralized data marketplace, it will be possible for data consumers and providers to interact and collaborate in Peer-to-Peer interactions \cite{park2018smart,ozyilmaz2018idmob}. DLTs enable peers to engage in financial transactions without establishing a trust relationship. Benefits often cited of DLTs, indeed, include enabling secure transactions between untrusted parties through consensus mechanisms, high availability, and the ability to automate and enforce processes through smart contracts \cite{buterin2013ethereum}.

With the management of market interactions based on the use of DLT and decentralized technologies, what remains is to manage the efficient search of data. Indeed, data inserted in the ledgers and DFS is usually unstructured and no efficient mechanisms are present to query about a certain kind of data. 
Thus, even if anyone can run public DLTs and DFS nodes, such as Ethereum~\cite{buterin2013ethereum} and IPFS~\cite{benet2014ipfs}, 
%to access to all the stored data, these ones are generally slow and expensive. 
data lookup can be very slow and expensive.
Data are rarely stored in a format that can be consumed directly and need to be filtered and indexed before any complex query. Data are referenced through addresses or indexes that, most of the time, are not related to the content of the data and are not useful for categorisation. 

In this paper we propose a system for the search of data according to their content or meaning. Our approach relies on the use 
%In this paper we propose the design and simulation 
of a Distributed Hash Table (DHT) as a layer placed over the DLTs. According to this solution, once acquired and recorded in DLTs, data can be searched through keywords thanks to the lookup features offered by the DHT. The distinctive feature of the DHT network is that it is essentially a hypercube overlay structure \cite{joung2007keyword}, in which each node will index objects representing specific indexed and addresses of a DLT using keywords. An interesting aspect of the specific hypercube-based DHT is that it allows to efficiently search for objects matching a specific keywords set $K$. Moreover, it allows searching for supersets of $K$, thus enabling the construction of queries which are more complex than a single $<$keyword, value$>$ lookup.

We take decentralized data markets as use case for our proposal, however, our approach is independent to the underlying DLT and can be easily extended to other distributed ledgers and DFS.
We provide a specific system design that is tailor-made for IOTA, a DLT designed for the IoT industry and data sharing, that exploits the use of a Direct Acyclical Graph (DAG), i.e. the Tangle \cite{popov2016tangle}. 
We also provide results coming from a detailed simulation analysis, which confirm that our system allows for multiple keyword searches in reasonable time 
(of the order of the logarithm of the hypercube nodes number).
%(of the order of the logarithm of the entire keyword set cardinality). 
We thus give a first contribution towards the creation of a system which allows for complex queries on top of decentralized ledgers and file systems.

The remainder of the paper is organized as follows. Section \ref{sec:back} provides a background on the technologies used. Section \ref{sec:arch} present a description of the architecture and of the DHT structure. In Section \ref{sec:sim} the system simulation is showed with some results that are discussed in \ref{sec:disc}. Finally, Section \ref{sec:concl} provides the concluding remarks.

%%%%%%%%%%%%%%%%%%%%%%%%%%%%%%%%%%%%%%%%%%%%%%%%%%%%%%%%%%%%%%%%%%%%%%%%%%%%%%%%%%%%%%%%%%%%%%%%%%%%%%%%%%%%%%%%%%%%%%%%%%%%%%%%%%%%%%%%%%%%%%%%%%%%%%%%%%%%%%%%%%%%%%%%%%%%%%%%%%%%%%%%%%%%%%%%%%%%%%%%%%%%%%%%%%%%%%%%%%%%%%%%%
\section{Background}\label{sec:back}
\subsection{Distributed Hash Table (DHT)}
A Distributed Hash Table (DHT) is a distributed infrastructure and storage system that provides the functionalities of a hash table, i.e. a data structure that efficiently maps ``keys'' into ``values''.
It consists on a peer-to-peer network of nodes that are supplied with the table data and on a routing mechanism that allows searching for objects in the network~\cite{joung2007keyword}. Each node in the DHT network is responsible for part of the entire system's keys and allows the objects mapped to the keys to be reached. 
In addition, each node stores a partial view of the entire network, with which it communicates for routing information. To reach nodes from one part of the network to another, a routing procedure typically traverses several nodes, approaching the destination at each hop.
This type of infrastructure has been used as a key element to implement complex and decentralized services, such as Content-Addressable Networks (CANs) \cite{ratnasamy2001scalable}, DFS \cite{benet2014ipfs}, cooperative web caching, multicast and domain name services. 

\subsection{Distributed Ledger Technologies (DLTs)}
A Distributed Ledger Technology (DLT) consists in a network of nodes, each of which maintains a replicated copy of a data ledger. The updates to the ledger are agreed following a consensus mechanism. 
%These kinds of technologies are thought with the aim to move trust from a human intermediary, that manages a transaction between two parties, to their protocol. 
Consensus mechanisms are implemented in order to enable two parties to transact directly without the need of a third party.
The main peculiarity of DLTs is that they ensure untampered data availability. Thus, they promote the development of trustful and reliable service applications \cite{aiello2020ippo,onik2019privacy,ozyilmaz2018idmob,zichichi2020framework}.

There are different implementations of DLTs, each one with its pros and cons. 
%In permissionless ones, anyone can take part to the consensus mechanism, while this is not true in permissioned ones. Another 
One of the main distinctions lies on the support of smart contracts, e.g. Ethereum~\cite{buterin2013ethereum}. This feature is quite often in contrast with other key features, related to the level of scalability and responsiveness of the system \cite{bez2019scalability}. Conversely, some implementations are thought to provide better scalability at the expense of lacking some features, e.g. based on Direct Acyclical Graphs (DAGs).

\subsubsection{IOTA}
IOTA is a DLT that allows hosts in a network to transfer immutable data among each other. In the IOTA ledger, i.e. the Tangle~\cite{popov2016tangle}, the vertices of a DAG represent transactions and edges represent validations to previous transactions. 
%When a new transaction is to be issued, two previous transactions must be selected (i.e. tips selection) and approved by referencing those in the transaction. The result is represented by means of directed edges in the Tangle. To validate a transaction a Proof-of-Work is performed (in order to deter denial of service attacks and other service abuses).
The validation approach is thought to address two major issues of traditional blockchain-based DLTs, i.e.~latency and fees. IOTA has been designed to offer fast validation, and no fees are required to add a transaction to the Tangle \cite{BROGAN2018257}. 

An important feature offered by IOTA is the Masked Authenticated Messaging (MAM). 
%MAM is a layer two data communication protocol built upon the Tangle, which adds the functionality to emit and access an encrypted data stream over the Tangle~\cite{BROGAN2018257}. 
MAM is a data communication protocol built upon the Tangle, which adds the functionality to emit and access an encrypted data stream over the Tangle~\cite{BROGAN2018257}. Since the MAM protocol relies over the underlying Tangle, it is referred as a "layer two" solution.
Data streams assume the form of channels, i.e.~a linked list of ordered messages stored in transactions. Once a channel is created, only the channel owner can publish encrypted messages on it. Users that possess the MAM channel encryption key (or set of keys, since each message can be encrypted using a different key) are enabled to decode the message and messages are addressed by a ``root'' value. 
%Messages are pushed on the channel in chronological order, and each message has a link to the next message to be created. Thus, once a user gains access to the MAM channel, he is enabled to see data from that moment on, whilst he cannot look back through the history of the channel before his entrance. 
In other words, MAM enables users to subscribe and follow a stream of data, generated by some device. 
From a functional point of view, channels are an ordered set of messages, in fact a channel is referenced through the root of a ``starting'' message. 

\subsection{Related Works}
The popularity of IoT devices and smartphones and the associated generation of large amounts of data derived from their sensors has resulted in an interest of individuals in the production and consumption of data via a data marketplace \cite{crabtree2018building}. Making data (which are mostly personal) available for access and trade is expected to become a part of the data-driven digital economy \cite{euc-com767}. As introduced earlier, the use of DLTs has been proposed for the implementation of data marketplaces to take advantage of \cite{de2018peer,zhu2020keyword}: (i) no need to rely on third party platforms; (ii) better resilience against network partitioning and single point of failure; (iii) privacy preserving mechanisms~\cite{onik2019privacy}. 
Most of the related works investigate on the data distribution through DLTs, focusing in particular on the use of off-chain storage based on DFS with data links referenced in DLTs \cite{aiello2020ippo,onik2019privacy,zichichi2020framework}.
In \cite{8656952}, authors provide the implementation of a data marketplace based on the use of DFS for storing data and a payment protocol that exploits Ethereum smart contracts \cite{buterin2013ethereum}.
Similarly, in \cite{park2018smart,ozyilmaz2018idmob} the proposed systems are based on P2P interactions and smart contracts to reach an agreement, while also integrating other components such as the IOTA DLT.

On the other hand, decentralized data search on DLT and DFS is a broader field that has been addressed by both scholars and developers. The Graph is one of the first protocols with the aim to provide a ``Decentralized Query Protocol''~\cite{thegraph2020protocol}. The Graph network consists in a layer two protocol based on the use of a Service Addressable Network, i.e. a P2P network for locating nodes capable of providing a particular service such as computational work (instead of objects just as a CAN).
In \cite{JIANG2020781}, authors propose a layer one keyword search scheme which implements oblivious keyword search in DFS. Their protocol is based on keywords search with authorization for maintaining privacy with retrieval requests stored as a transaction in a blockchain (i.e. layer one).
Finally, a layer two solution for keywords search in DFS has been proposed in \cite{electronics9122041}, where a combination of a decentralized B+Tree and HashMaps is used to index IPFS objects \cite{benet2014ipfs}.

%%%%%%%%%%%%%%%%%%%%%%%%%%%%%%%%%%%%%%%%%%%%%%%%%%%%%%%%%%%%%%%%%%%%%%%%%%%%%%%%%%%%%%%%%%%%%%%%%%%%%%%%%%%%%%%%%%%%%%%%%%%%%%%%%%%%%%%%%%%%%%%%%%%%%%%%%%%%%%%%%%%%%%%%%%%%%%%%%%%%%%%%%%%%%%%%%%%%%%%%%%%%%%%%%%%%%%%%%%%%%%%%%
\section{System Architecture}\label{sec:arch}
\subsection{Decentralized Data Marketplace}
\begin{figure}
    \centering
    \includegraphics[width=.49\textwidth]{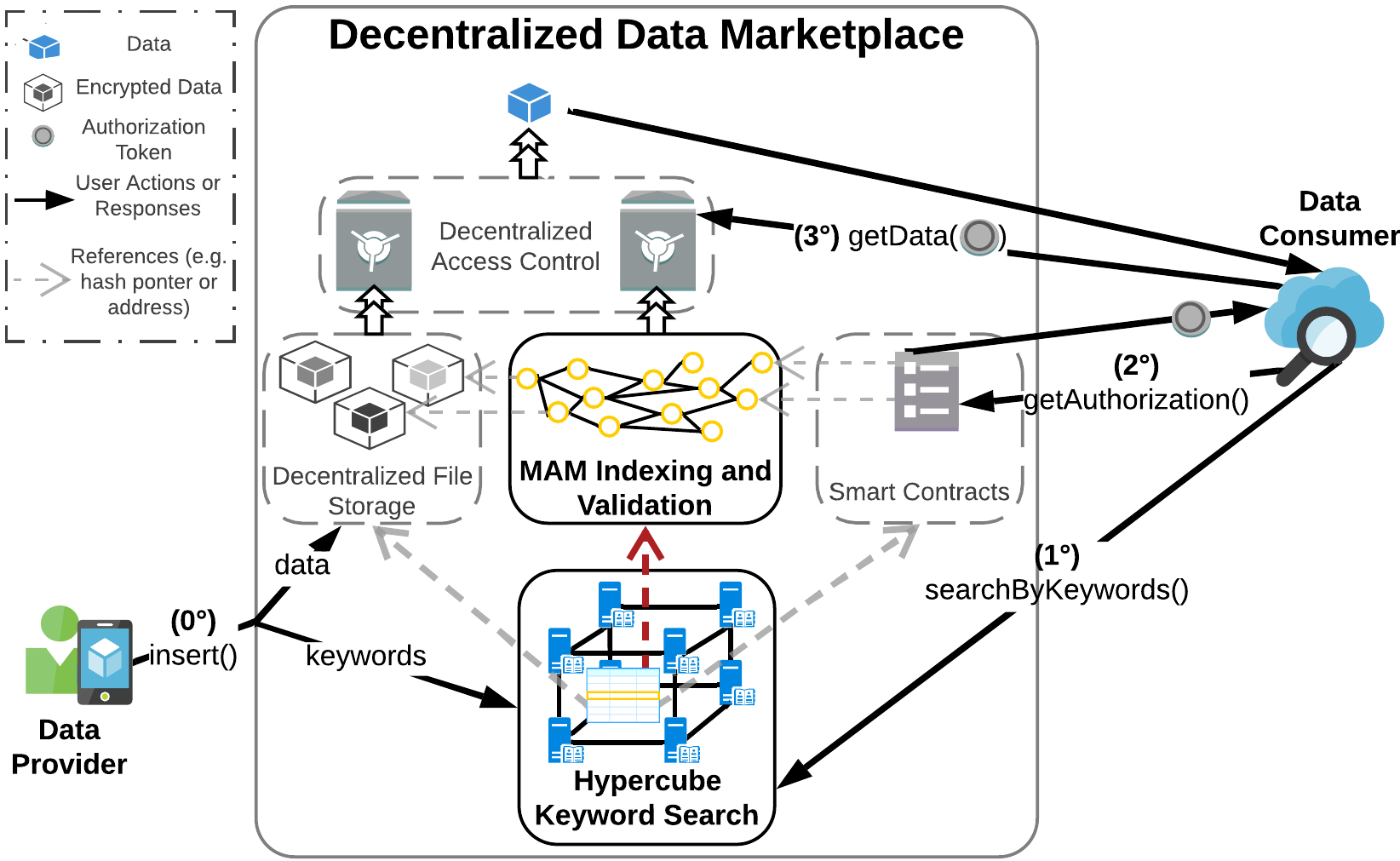}
    \caption{Decentralized Data Marketplace architecture scheme.}
    \label{fig:market}
\end{figure}

Our contribution focuses on the design of a system that allows to create queries based on the multiple keyword searches. We apply our solution to the specific context of a decentralized data market. 
%However, the solution we provide can be leveraged in several cases, especially when decentralization is required. 
As shown in Figure \ref{fig:market}, the different architectural components are arranged to provide a data marketplace service, based on decentralized technologies:
\begin{itemize}
    \item Decentralized File Storages (DFS) can be used to store data in an encrypted form, offering high availability \cite{zichichi2020efficiency}. 
    \item A decentralized access control system can be leveraged by data consumers to get the data from the DFS once they have been authorized (e.g. through a token) \cite{zichichi2020personal}.  
    \item Smart contracts have the ability to provide a distributed authorization mechanism following a policy indicated by the data provider (e.g. access through payment) \cite{zichichi2020framework}. 
    \item A DLT such as IOTA and the use of MAM channels enable the data indexing and validation (in form of hash pointers) \cite{zichichi2020are,zichichi2020framework}.
    \item A distributed mechanism for the search of data is in charge of associating keywords to addresses or references stored in DLTs, smart contracts and DFS.
\end{itemize}

In this paper, we specifically focus on the retrieval of data stored in DLTs (red dashed arrow in Figure \ref{fig:market}). In particular, our use case consists in the research of data stored in IOTA MAM channels messages. These messages contain data itself (in an encrypted form) or a reference to the data stored in DFS, e.g. hash pointers.

\subsection{Layer Two Lookup Scheme}
\begin{figure}
    \centering
    \includegraphics[width=.35\textwidth]{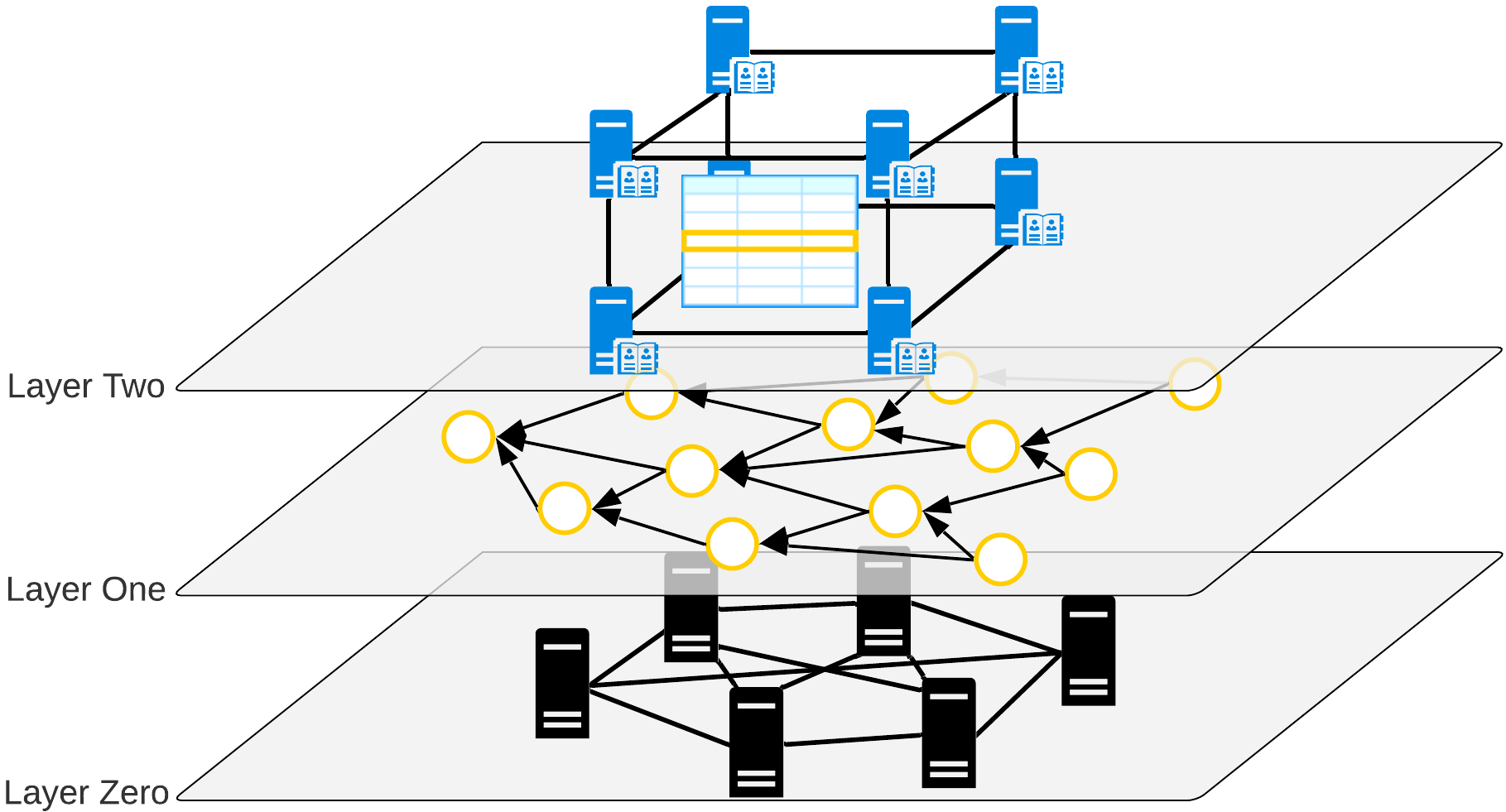}
    \caption{Layers in the context of DLTs. Layer zero consists in the DLT network, while layer one is the set of software frameworks run by the network nodes (e.g. the ledger). Layer two solutions are the ones that leverage layer one for other services, i.e. the DHT Keyword Search in our case.}
    \label{fig:layers}
\end{figure}
We designed a layer two solution using a DHT with the aim of facilitating the search of large amounts of data through specific keywords (Figure \ref{fig:layers}). 
%There are several implementations of DLT that vary in the form of the ledger (e.g. blockchain, DAG), its distribution (e.g. sharding), consensus mechanism (Proof-of-Work, Proof-of-Stake) and features (e.g. smart contracts). 
While layer one protocols and technologies in DLTs define the form of the ledger, its distribution, consensus mechanism and features, layer two solutions are built on top of layer one without changing its trust assumptions, i.e. the consensus mechanism, or the structure\cite{gudgeon2020sok}. Layer two protocols allow users to communicate through private channels, reducing the transaction load on the underlying DLT.

One of the main use case of DLTs consists in data sharing, e.g. vehicles \cite{zichichi2020framework,zichichi2020are}, IoT devices \cite{crabtree2018building}, smartphone data and personal data \cite{onik2019privacy,zichichi2020personal}. Once collected, data are stored in a DFS and/or referenced in a DLT via hash digests. In our case, we use the IOTA DLT because it allows us to manage the upload of data in the form of messages streams thanks to the MAM protocol. 
%As already mentioned in Section \ref{sec:back}, the MAM protocol allows the creation and subscription to channels that contain data streams in the form of messages. 

In order to obtain information from a message within a MAM channel, it is necessary to know the exact address of the message or of the channel, i.e. the root value. 
%However, knowing only the root of a message is very limiting to obtain any information, or else it is not possible to know all the roots of a given search topic. 
However, the root of a MAM channel does not provide any information related to the type and kind of messages. No mechanisms are provided for the discovery based on the content of certain data/MAM channels that are available in the Tangle. This is the issue we are dealing with in this paper.
To surmount such limitation, in our system every single message included in a MAM channel is indexed by a keyword set. Such keyword set is then exploited to search for specific kinds of contents.

\subsection{Hypercube based DHT}
Considering $O$ as the set of all MAM messages in IOTA, the idea is to map each object $o \in O$ to a keyword set $K_o \subseteq W$, where $W$ is the keyword space, i.e. the set of all keywords considered. In general, we refer to $K \subseteq W$ as a keyword set, that can be associated to a data content (i.e. the metadata associated to it) or a query (i.e. we are looking to some content with a specific metadata). By using a uniform hash function $h : W \rightarrow \{0, 1, \hdots,r-1\}$, a keyword set $K$ can be represented by the result of such function, i.e. a string of bits $u$ where the $1$s are set in the positions given by $\textit{one}(u) = \{ h(k) \mid k \in K \}$. In other words, each $k \in W$ has a fixed position in the $r$-bit string given by $h(k)$ and that position can be associated to more than one $k$ (i.e. hash collision). Then, every keyword set $K$ is represented by a $r$-bit string where the positions are ``activated'', i.e. are set to 1, by all the $k \in K$. 

We use these $r$-bit strings to identify logical nodes in a DHT network, e.g. for $r=4$ a node id can take values such as $0100$ or $1110$.
In particular, inspired to \cite{joung2007keyword}, we refer to the geometric form of the hypercube to organise the topological structure of such DHT network. $H_r(V,E)$ is a $r$-dimensional hypercube, with a set of vertices $V$ and a set of edges $E$ connecting them. Each of the $2^r$ vertices represents a logical node, whilst, edges are formed when two vertices differ of only one bit, e.g. $1011$ and $1010$ share an edge. In the network, the nodes represented by vertices that share an edge are network neighbors as well.
To find out how far apart two vertices $u$ and $v$ are within the hypercube, the Hamming distance can be used, i.e. $\textit{Hamming}(u,v)= \sum_{i=0}^{r-1}(u_i \oplus v_i),$ where $\oplus$ is the XOR operation and $u_i$ is the bit at the $i$-th position of the $u$ string, e.g. for $u=1011$ and $v=1010$, we have $\textit{Hamming}(u,v)=1$.

\subsection{Keyword-based Complex Queries}
In our system, contents can be discovered through queries that are based on the lookup of multiple keywords, associated to data. Such queries are processed by the DHT-based indexing scheme described in the previous section.
The base idea is to associate a keyword set to each MAM message through the DHT. In particular each logical node will locally store an index table that associates a keyword set $K_o$ to the root of a MAM message, i.e. the reference of an object $o$.  
Then, given a keyword set $K$, the associated $r$-bit string is used to reach the logical node responsible for $K$ through a routing mechanism, in order to obtain the set of $\text{objects}=\{o \in O \mid K_o \supseteq K\}$. 
For instance, with $W =\{$\textit{``Bologna'', ``San Donato'', ``Temperature'', ``Celsius''}$\}$ and $1010$ representing the keyword set $K=\{$\textit{``Bologna, Temperature''}$\}$, if $u \in V$ is the node which is responsible for $K$ because the id of $u$ is equal to $1010$, then $u$ is in charge of maintaining a list of roots of MAM messages containing the temperature of the city of Bologna. 
%Once that node is located, the $\text{objects}=\{o \in O \mid K_o = K\}$ it stores in its index table can be returned or aggregated with other nodes' objects (depending on the search types showed in the next sub-section). These objects consist of a list of roots that can be used to get MAM messages from the Tangle.
 
\subsubsection{Multiple Keywords Search}
Our system provides two functions for making queries based on multiple keywords:
\begin{itemize}
    \item \textbf{Pin Search} - this procedure aims at obtaining all and only the objects associated exactly with a keyword set $K$, i.e., $\{o \in O \mid K_o = K\}$. Upon request, the responsible node returns to the requester all the roots of the corresponding objects that it keeps in its table, associated to $K$.
    \item \textbf{Superset Search} - this procedure is similar to the previous one, but in addition it also searches for objects that can be described by keywords sets that include K, i.e., $\{o \in O \mid K_o \supseteq K\}$. Since the possible outcomes of this search can be quite large, a limit $l$ is set. 
\end{itemize}
For the Pin Search we need to retrieve objects only from one node. Whilst, for Superset Search, we need to retrieve objects from all nodes that are responsible for a Superset of $K$. Such nodes are contained in the sub-hypercube $SH(S,F)$ induced by the node $u$ responsible for $K$, where $S$ includes all the nodes $s \in V$ that ``contain'' $u$, i.e., $u_i = 1 \Rightarrow w_i = 1$, while $F$ includes all the edges $e \in E$ between such nodes.
Thus, during a Superset Search, the induced sub-hypercube is computed and then only nodes in such sub-hypercube are queried using a spanning binomial tree as described in \cite{joung2007keyword} (definition 4.2). The $l$ limit is a query parameter that indicates the maximum amount of objects to return when traversing the spanning binomial tree.

\subsubsection{The Query Routing Mechanism}
Queries can be injected into the system by users external to the DHT to any $v \in V$ network node. Through a routing mechanism, the query will reach reach a node $u \in V$ that is responsible for a keyword set $K$. This process is described in detail in Algorithm \ref{algo}.
%Then considering node  as the responsible for the keyword set $K$, the routing mechanism from a node $v \in V$ to $u$ with a keyword set $K_o$ is as follows:
%\begin{enumerate}
    %\item node $v$ computes the Hamming distance to node $u$ for all its %neighbor nodes;
    %\item node $v$ broadcasts the query to the neighbor $w$ with the lowest %distance to $u$;
    %\item this process is repeated by $w$ until the query reaches $u$;
    %\item if the query is a Pin Search $u$ returns the objects references %associated to $K$;
    %\item else (in the case of a Superset Search) $v$ computes its children %in the spanning binomial tree of the induced sub-hypercube;
    %\item node $u$ broadcasts the query to the children;
    %\item the children will repeat the process from step 5 with their %children and then return the objects when the limit $l$ is reached.
%\end{enumerate}
\begin{algorithm}
    \caption{QueryRoutingMechanism}
    \label{algo}
    \SetAlgoLined
    \DontPrintSemicolon
    \KwIn{$q$ query, $K$ keyword set, $l$ limit}
    \KwData{$v$ node string, $\textit{one}(v)$, $\textit{neighbors}(v)$}
    \KwResult{$\{o \in O \mid K_o \supseteq K\} $}
    %\;
    %\SetKwFunction{FMain}{publishData}%
    %\SetKwProg{Fn}{Function}{:}{\KwRet}
    %\Fn{\FMain{$p$, $id_{FC}$}}{
    $\textit{one}(u) \leftarrow \{ h(k) \mid k \in K \}$\;
    \eIf{$\textit{one}(u) \neq \textit{one}(v) \wedge \text{From}(q) = ``\textit{User}``$}{
        $\textit{w} \leftarrow \{ n \mid n \in \textit{neighbors}(v) \wedge$ $\text{Min}(\text{Hamming}(n,u)) \}$\;
        \KwRet QueryRoutingMechanism($w, q, K, l$)\;
    }{
        \uIf{$\text{Type}(q) = ``\textit{PinSearch}``$}{
            \KwRet GetObjectsFromIndexTable($K$, $-1$)\;
        }
        \ElseIf( //\textit{ i.e. SupersetSearch}){$\textit{one}(u) \subseteq \textit{one}(v)$}{
        $\textit{objectsList} \leftarrow$ GetObjectsFromIndexTable($K$, $l$)\;
        $l \leftarrow l - $ Length(\textit{objectsList})\;
        $\text{From}(q) \leftarrow ``\textit{Node}``$\;
        \While{$l > 0$}{
            $\textit{c} \leftarrow$ $\text{GetNextSBTreeChild}(u)$\;
            $\textit{cList} \leftarrow$ QueryRoutingMechanism($c, q, K, l$)\;
            \textit{objectsList} $\leftarrow$ \textit{objectsList} + \textit{cList} \;
            $l \leftarrow l - $ Length(\textit{cList})\;
        }
        \KwRet objectsList \;
        }
    }
\end{algorithm}
%%%%%%%%%%%%%%%%%%%%%%%%%%%%%%%%%%%%%%%%%%%%%%%%%%%%%%%%%%%%%%%%%%%%%%%%%%%%%%%%%%%%%%%%%%%%%%%%%%%%%%%%%%%%%%%%%%%%%%%%%%%%%%%%%%%%%%%%%%%%%%%%%%%%%%%%%%%%%%%%%%%%%%%%%%%%%%%%%%%%%%%%%%%%%%%%%%%%%%%%%%%%%%%%%%%%%%%%%%%%%%%%%
\section{Performance Evaluation}\label{sec:sim}
\begin{figure*}[ht]
    \centering
    \includegraphics[width=.85\textwidth]{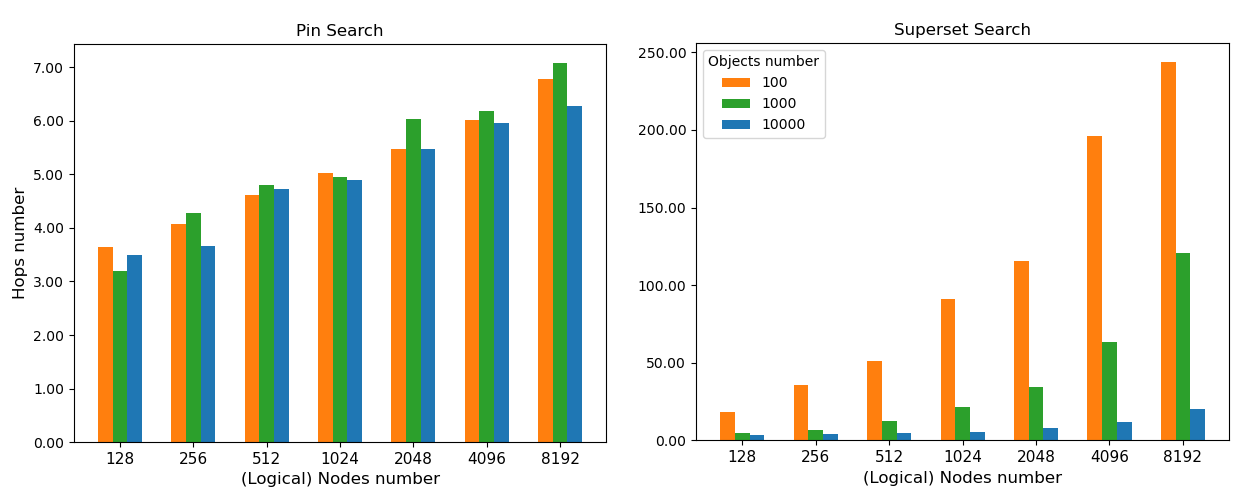}
    \caption{Number of hops on average for the Pin Search (left) and Superset Search (right).}
    \label{fig:pin}
\end{figure*}

\begin{table*}
\caption{Pin Search Number of Hops.}
\begin{center}
\begin{tabular}{ |c|c|c|c|c|c|c|c|c|c| } 
\hline
\textbf{Nodes Num.} & \multicolumn{3}{c|}{\textbf{Average}} & \multicolumn{3}{c|}{\textbf{Standard Deviation}} & \multicolumn{3}{c|}{\textbf{Confidence Interval (95\%)}}\\
\cline{2-10} & 100 & 1000 & 10000 & 100 & 1000 & 10000 & 100 & 1000 & 10000 \\
\hline \hline
\textbf{128} & 3.64 & 3.2 & 3.5 & 1.33 & 1.32 & 1.12 & (3.27,4.01) & (2.83,3.57) & (3.19,3.81) \\ 
\hline
\textbf{256} & 4.08 & 4.28 & 3.66 & 1.45 & 1.48 & 1.31 & (3.68,4.48) & (3.87,4.69) & (3.29,4.03) \\ 
\hline
\textbf{512} & 4.62 & 4.8 & 4.72 & 1.57 & 1.70 & 1.24 & (4.18,5.06) & (4.33,5.27) & (4.37,5.07) \\ 
\hline
\textbf{1024} & 5.02 & 4.96 & 4.9 & 1.68 & 1.67 & 1.69 & (4.55,5.49) & (4.49,5.43) & (4.43,5.37) \\ 
\hline
\textbf{2048} & 5.48 & 6.04 & 5.48 & 1.76 & 1.85 & 1.69 & (4.99,5.97) & (5.53,6.55) & (5.01,5.95) \\ 
\hline
\textbf{4096} & 6.02 & 6.18 & 5.96 & 1.55 & 1.61 & 1.62 & (5.59,6.45) & (5.73,6.63) & (5.51,6.41) \\ 
\hline
\textbf{8192} & 6.78 & 7.08 & 6.28 & 1.63 & 1.60 & 1.64 & (6.33,7.23) & (6.64,7.52) & (5.82,6.74) \\ 
\hline
\end{tabular}
\end{center}
\label{fig:table}
\end{table*}

\begin{table*}
\caption{Superset Search Number of Hops.}
\begin{center}
\begin{tabular}{ |c|c|c|c|c|c|c|c|c|c| } 
\hline
\textbf{Nodes Num.} & \multicolumn{3}{c|}{\textbf{Average}} & \multicolumn{3}{c|}{\textbf{Standard Deviation}} & \multicolumn{3}{c|}{\textbf{Confidence Interval (95\%)}}\\
\cline{2-10} & 100 & 1000 & 10000 & 100 & 1000 & 10000 & 100 & 1000 & 10000 \\
\hline \hline
\textbf{128} & 18.28 & 4.54 & 3.52 & 8.44 & 1.54 & 1.19 & (15.94,20.62) & (4.11,4.97) & (3.19,3.85) \\
\hline
\textbf{256} & 35.90  & 6.80 & 4.16 & 17.89 & 2.25 & 1.43 & (30.94,40.86) & (6.17,7.43) & (3.76,4.56) \\
\hline
\textbf{512} & 51.18 & 12.16 & 4.46 & 37.85 & 3.29 & 1.31 & (40.69,61.67) & (11.25,13.07) & (4.10,4.82) \\
\hline
\textbf{1024} & 91.06 & 21.70  & 5.08 & 72.44 & 6.23 & 1.68 & (70.98,111.14) & (19.97,23.43) & (4.61,5.55) \\
\hline
\textbf{2048} & 115.70 & 34.56 & 7.84 & 98.39 & 13.00 & 1.98 & (88.43,142.97) & (30.96,38.16) & (7.29,8.39) \\
\hline
\textbf{4096} & 196.00 & 63.38 & 11.92 & 186.88 & 25.37 & 2.64 & (144.20,247.80) & (56.35,70.41) & (11.19,12.65) \\ 
\hline
\textbf{8192} & 243.90 & 120.38 & 20.38 & 253.59 & 68.65 & 6.28 & (173.61,314.19) & (101.35,139.41) & (18.64,22.12) \\
\hline
\end{tabular}
\end{center}
\label{fig:table2}
\end{table*}

We conducted a simulation assessment using PeerSim, a simulation environment developed to build P2P networks using extensible and pluggable components \cite{montresor2009peersim}. Once designed and implemented the hypercube structured DHT for multiple keyword search, we focused on the study of the efficiency of the routing mechanism. Several tests were carried out assuming different scenarios 
%in which the network was composed of a variable number of nodes and stored objects. I
and in the following, we report on the main outcomes we obtained.

\subsection{Evaluation}
\subsubsection{Tests Setup}
In order to evaluate Pin Search and Superset Search, tests were carried out on different sizes of the hypercube. Specifically, the number of nodes was varied from 128 ($r = 7$) up to 8192 ($r = 13$).  Then, for each dimension $r$ a different number of randomly created keywords-objects (i.e. MAM message roots) was inserted in the DHT. The number of objects taken into consideration varies from $100$, $1000$ and finally $10000$.
Given the nature of the tests, i.e., a simulated network, we considered the number of hops required for each new query as a parameter to be evaluated. The query keyword sets were randomly generated and the starting node randomly chosen.
For each type of test, $50$ repetitions were performed, and then the average results were calculated. For the Superset search, the limit value was set to $l=10$ objects.

\subsubsection{Pin Search}
As shown in Figure \ref{fig:pin} (left), the number of hops required to transmit a message from the source node to the destination node increases as the hypercube dimension increases, i.e. nodes number. The average number of hops increases from about $3.5$ for $128$ nodes ($r = 7$) to about $6.72$ for $8192$ nodes ($r = 13$).
This behavior can be explained by the fact that by increasing the hypercube dimension the path that a message must take before reaching its destination is automatically enlarged.
The number of objects in the testbed does not affects the final outcome, since the path to reach the target node only follows the rationale of the hypercube and does not depend on the number of keyword-object associations stored in the DHT.

\subsubsection{Superset Search}
The tests performed on the Superset Search present results with dissimilar values with respect to the previous case (Figure \ref{fig:pin}, right). 
At a first glance, in fact, those apparently anomalous values stand out, corresponding to a high number of hops between nodes, which decreases with the referenced objects number.
%With a low number of objects referenced in the DHT, there are high average number of hops needed to satisfy the Superset search.
This phenomenon can be explained by the fact that the Superset search traverses the spanning binomial tree of the sub-hypercube induced by the node responsible for the keyword set, until it finds the number of objects indicated by the limit, i.e. $l=10$.
Hence, in a network with many nodes and few objects, the query might take longer to reach that limit, because many nodes are ``empty'', i.e. do not reference any object.
Considering the case of $4096$ nodes ($r = 12$) and $10000$ objects, 
in a Pin search $5.96$ hops are required, on average. In a Superset search other $11.92 - 5.96 = 5.96$ hops are needed to reach other nodes containing other results of the superset search, until the limit $l$ is reached.
%on average in the Superset Search, i.e. traversing the spanning binomial tree for a keyword set query until the limit $l$ is reached.
%If objects were uniformly distributed, the total number of nodes requested to return objects would have dropped to $4$ nodes because each node would have maintained $\frac{10000}{4096} = 2.44$ object references on average and $l=10 (\cong 4 \times 2.44)$.

%%%%%%%%%%%%%%%%%%%%%%%%%%%%%%%%%%%%%%%%%%%%%%%%%%%%%%%%%%%%%%%%%%%%%%%%%%%%%%%%%%%%%%%%%%%%%%%%%%%%%%%%%%%%%%%%%%%%%%%%%%%%%%%%%%%%%%%%%%%%%%%%%%%%%%%%%%%%%%%%%%%%%%%%%%%%%%%%%%%%%%%%%%%%%%%%%%%%%%%%%%%%%%%%%%%%%%%%%%%%%%%%%
\section{Discussion}\label{sec:disc}
The results provided in the previous section confirm what was expected due to the hypercube structure of the network: the Pin Search number of hops are of the order of the logarithm of the hypercube logical nodes number, i.e. $r$. In particular on average they are equal to $\frac{\log(n)}{2} = \frac{r}{2}$. For what concerns the Superset Search number of hops, on average, it is equal to $\frac{\log(n)}{2} + l $, where $l$ is the limit of the number of nodes in the sub-hypercube to reach. 

These results show the goodness of the solution in the trade-off between memory space and response time. In traditional DLTs, such as Ethereum and IOTA, searching for a datum in a transaction means traversing all the ``transaction sea'' in the ledger and for this reason the current solution is to use centralized ``DLT explorers''~\cite{explorer2020}. 
On the other hand, in the case of sharded DLTs, the proposed solution could become a layer one protocol to search the data between many shards.
  
Finally, while in this study we focused on DLTs as the underlying data storage, it is worth mentioning that, due to the origins of the hypercube proposal. \cite{joung2007keyword}, DFS systems can perfectly fit with such architecture, since most of them are based on DHT, already. Indeed, the implementation of the hypercube for keyword search in IPFS is matter of future work.

%%%%%%%%%%%%%%%%%%%%%%%%%%%%%%%%%%%%%%%%%%%%%%%%%%%%%%%%%%%%%%%%%%%%%%%%%%%%%%%%%%%%%%%%%%%%%%%%%%%%%%%%%%%%%%%%%%%%%%%%%%%%%%%%%%%%%%%%%%%%%%%%%%%%%%%%%%%%%%%%%%%%%%%%%%%%%%%%%%%%%%%%%%%%%%%%%%%%%%%%%%%%%%%%%%%%%%%%%%%%%%%%%
\section{Conclusions}\label{sec:concl}
In this paper, we have taken decentralized data markets as a use case and provided a layer two solution based on the use of DHT with the aim of facilitating the retrieval of large amounts of data using specific keywords. We focused specifically on retrieving data stored in IOTA MAM channel messages. However, our approach can be easily extended to other DLTs and DFSs.
The solution we provided consists of a DHT network structured as a hypercube to provide an efficient routing mechanism based on keyword sets.
We also provided some results from a detailed simulation analysis, which shows that searching for an object with an exact keyword set requires on average $\frac{\log(n)}{2}$ hops, where $n$ is the number of logical nodes of the hypercube.
This solution presents an efficient trade-off between memory space and response time, thus making a first contribution towards the creation of a system that allows complex queries on DLT and DFS.

%%%%%%%%%%%%%%%%%%%%%%%%%%%%%%%%%%%%%%%%%%%%%%%%%%%%%%%%%%%%%%%%%%%%%%%%%%%%%%%%%%%%%%%%%%%%%%%%%%%%%%%%%%%%%%%%%%%%%%%%%%%%%%%%%%%%%%%%%%%%%%%%%%%%%%%%%%%%%%%%%%%%%%%%%%%%%%%%%%%%%%%%%%%%%%%%%%%%%%%%%%%%%%%%%%%%%%%%%%%%%%%%%
\bibliographystyle{./IEEEtran}
\bibliography{./paper}

\end{document}